\documentclass{aa}

\usepackage{graphicx}
\usepackage{caption}
\usepackage{subcaption}
\usepackage{amssymb}
\usepackage{amsfonts}
\usepackage[export]{adjustbox}
\usepackage[varg]{txfonts}
\usepackage[utf8]{inputenc}
\usepackage{amsmath}
\usepackage{xcolor}
\usepackage{hyperref}
\usepackage{comment}
\usepackage[normalem]{ulem}
\usepackage{upgreek}

\newcommand{\rstar}{\ensuremath{R_{*}}}

\newcommand{\Msun}{\ensuremath{M_{\odot}}}
\newcommand{\Msunyr}{\ensuremath{M_{\odot}\, {\rm yr}^{-1}}}

\newcommand{\diff}{\ensuremath{{\rm d}}}

\newcommand{\epsBS}{\ensuremath{\epsilon_{\rm BS}}}
\newcommand{\betaBS}{\ensuremath{\beta_{\rm BS}}}
\newcommand{\conus}{\ensuremath{\theta_{\rm w}}}
\newcommand{\incl}{\ensuremath{i}}
\newcommand{\muang}{\ensuremath{\chi}}
\newcommand{\ximag}{\ensuremath{\xi_{\rm mag}}}
\newcommand{\keV}{\ensuremath{\rm keV}}
\newcommand{\cm}{\ensuremath{\rm cm}}
\newcommand{\Gcmc}{\ensuremath{\rm \,G \, cm^3}}
\newcommand{\ergl}{\ensuremath{\rm \,erg \, s^{-1}}}
\newcommand{\e}{\ensuremath{{\rm e}}}
\newcommand{\cs}{\ensuremath{c_{\rm s}}}
\newcommand{\Mach}{\ensuremath{{\cal M}}}
\newcommand{\thetamag}{\ensuremath{{\theta}}}

\renewcommand{\Re}{\ensuremath{R_{\rm e}}}
\newcommand{\Rm}{\ensuremath{R_{\rm m}}}
\newcommand{\mdot}{\ensuremath{{\dot M}}}

\newcommand{\EllipticE}[2]{\ensuremath{{\rm E}\, \left( #1\left| #2\right.\right)}}
\newcommand{\EllipticF}[2]{\ensuremath{{\rm F}\, \left( #1\left| #2\right.\right)}}

\newcommand{\fidu}{{\tt S}\xspace}
\newcommand{\zeroloss}{{\tt N}\xspace}

\newif\ifcleanversion

\cleanversionfalse 

\ifcleanversion

    \newcommand{\glnote}[1]{}
    \newcommand{\panote}[1]{}
    
    \newcommand{\del}[1]{}

    \newcommand{\glaccept}[1]{#1}
    
\else
    
    \newcommand{\glnote}[1]{\textit{\textcolor{blue}{#1}}}
    
    \newcommand{\panote}[1]{\textit{\textcolor{orange}{#1}}}

    \newcommand{\glaccept}[1]{{#1}\xspace}
    \newcommand{\del}[1]{\sout{#1}\xspace}
\fi

\defcitealias{BS76}{BS}
\defcitealias{AL23}{AL}
\defcitealias{variable-opacity}{LAII}

\begin{document}

\title{Shock-free super-Eddington accretion onto magnetized neutron stars}

\titlerunning{Shock-free accretion columns}

\author{
G. Lipunova\inst{1,2}
\and
P. Abolmasov\inst{3}
}

\institute{
Dr.~Karl Remeis-Observatory and Erlangen Centre for Astroparticle Physics, Friedrich-Alexander Universit\"at Erlangen-N\"urnberg, Sternwartstr.~7, 96049 Bamberg, Germany; \email{gvlipunova@gmail.com}
\and
Max-Planck-Institut f\"ur Radioastronomie, Auf dem H\"ugel 69, 53121 Bonn, Germany
\and
The Raymond and Beverly Sackler School of Physics and Astronomy, Tel Aviv University, Tel Aviv 69978, Israel
}

\abstract
{Strong magnetic fields allow accreting neutron stars to form magnetically controlled funnel flows. At sufficiently high accretion rates, such flows become radiation-pressure-supported. In the conventional picture, the magnetospheric flow contains a nearly free-falling transonic outer region, a radiation-mediated shock, and a subsonic accretion column.}
{We investigate the regime in which the the height of the column is expected to exceed  the magnetospheric radius. In this case a self-consistent transonic solution with a standing shock is no longer possible, and the entire magnetospheric flow remains subsonic.}
{We construct a semi-analytical hydrodynamical solution for a fully subsonic, radiation-supported magnetospheric flow in the low-Mach-number limit. The magnetic field is treated in the force-free approximation, while the flow structure is determined by mass conservation, energy conservation, and the momentum equation along the field lines. We test the solution using one-dimensional time-dependent simulations with the code {\tt HACol}.}
{The shock-free solution is highly advective. Its enthalpy profile is close to the adiabatic limit, while the energy density is regulated by leakage of mass and heat through the sides of the accretion channel. Most of the observable power is expected to be released by plasma leaving the magnetospheric flow above the neutron-star surface. Such systems should be associated with super-Eddington accretion discs and low pulse fractions due to visibility selection and scattering in the disc wind. We suggest that many non-pulsating ultraluminous X-ray sources containing neutron stars may accrete in this shock-free regime.}
{}


\maketitle

\section{Introduction}

The observational properties of an accreting neutron star (NS) depend strongly on whether the stellar magnetic field is able to disrupt the accretion flow and form a magnetosphere. For an accretion rate close to the Eddington value, this requires a surface magnetic field of order $\gtrsim 10^8$~G. Magnetic fields redirect the flow, force the accreting matter to follow magnetic field lines, and couple the inner flow to the stellar rotation. As a result, the observed X-ray flux is often modulated at the spin period of the NS, and the source is identified as an X-ray pulsar \citep[see][]{2012MmSAI..83..230C,2023hxga.book..138M}.

Typical X-ray pulsars have surface magnetic fields $B\sim 10^{12}-10^{13}$~G and are frequently found in high-mass X-ray binaries. Their luminosities span a very broad range, from quiescent values around $10^{32}\ergl$ \citep{2020A&A...638A.152R} to super-Eddington luminosities above $10^{39}\ergl$ \citep{2018MNRAS.479L.134T}. The discovery of pulsating ultraluminous X-ray sources demonstrates that accretion onto magnetized NSs can proceed at rates tens to thousands of times higher than the classical Eddington value.

At low luminosities, matter reaches the stellar surface and forms a hot spot. At higher luminosities, radiation pressure becomes dynamically important. The infalling plasma is then decelerated above the NS surface by a radiation-mediated shock, and the region below the shock forms an accretion column. The classical theory of radiation-supported magnetospheric accretion was developed by \citet{BS76}, hereafter \citetalias{BS76}, following earlier work by \citet{Davidson1973} and \citet{Inoue1975}.

The structure of the accretion column depends on the ability of radiation to diffuse through the column walls. If cooling is efficient, the column remains relatively short and most of the released gravitational energy is radiated locally. If accretion rate exceeds a certain limit, the cooling of the column is inefficient, and a substantial fraction of the energy is advected downward with the flow (see the accompanying paper, \citealt{variable-opacity}, hereafter \citetalias{variable-opacity}). The latter regime is the sinking regime of \citetalias{BS76}. Time-dependent simulations by \citet{AL23}, hereafter \citetalias{AL23}, confirmed the existence of such advective columns and showed that, at sufficiently high accretion rates, mass and heat can escape through the sides of the accretion channel.

According to both the analytical solution of \citetalias{BS76} and the simulations of \citetalias{AL23}, sufficiently large accretion rates do not allow a steady radiative shock to remain inside the magnetosphere. One then expects the flow to switch to a different regime: a shock-free magnetospheric flow in which the plasma remains subsonic between the inner disc and the stellar surface. This regime is the focus of the present work.

Such flows are relevant for the interpretation of ultraluminous X-ray sources (ULXs). ULXs are off-nuclear sources with apparent luminosities $\gtrsim 10^{39}\ergl$ \citep{2017ARA&A..55..303K}. Some ULXs show coherent pulsations and therefore contain accreting NSs \citep{2014Natur.514..202B,2016ApJ...831L..14F,2017Sci...355..817I}. However, the fraction of ULXs with detected pulsations is small, of order one per cent \citep{2015A&A...579A..22D}. This raises the question of whether most ULXs contain NSs whose pulsations are suppressed by geometry, scattering, or accretion-regime effects.

The paper is organized as follows. In Sect.~\ref{sec:an}, we derive the semi-analytical shock-free solution. In Sect.~\ref{sec:sim}, we compare this solution with time-dependent numerical simulations. In Sect.~\ref{sec:obs}, we discuss the luminosity, spectrum, variability, population implications, and pulse fractions expected in this regime. We conclude in Sect.~\ref{sec:conc}. A detailed discussion of opacity-dependent transitions between accretion-column regimes is presented in a companion paper.

\section{Analytic shock-free solution}\label{sec:an}

\subsection{General system of equations}\label{sec:an:gen}

We consider a flow in the hydrodynamic ``bead-on-a-wire'' regime (see for example \citealt{1971MNRAS.152..323H}) that exists within a nearly force-free magnetosphere.
Momentum equation projected upon the field lines is purely hydrodynamical and may be written as 
\begin{equation}\label{E:Euler}
    v \frac{\diff v}{\diff l} =  g_\parallel - \frac{1}{\rho} \frac{\diff P}{\diff l}\, ,
\end{equation}
where $v$ is the infall velocity (directed downwards, we will assume $v>0$ throughout the paper), $P$ and $\rho$ are pressure and density, $g_\parallel$ is the total mass force (gravity + centrifugal) projected onto the field line, $l$ is the coordinate along the field line, that we will express (assuming aligned dipolar geometry) as
\begin{equation}
\label{E:dl}
        \diff l = - \Re \sqrt{1+3\cos^2\theta} ~ \diff\cos\theta\, .
\end{equation}
Here $\Re$ is the outer radius of the dipolar field line, which equals the  magnetospheric radius $\Rm$ for an aligned dipole, and $\theta$ is the polar angle of the spherical coordinate system. 
The total mass force in presence of rigid-body rotation at a frequency of $\Omega$ is (according to \citetalias[equations 27 and A10-A11]{AL23})
\begin{equation}
\begin{array}{l}
\displaystyle    g_\parallel = - \frac{2\cos\theta}{\sqrt{1+3\cos^2\theta}} \frac{GM}{R^2} + \Omega^2 R \frac{3\cos\theta \sin^2\theta}{\sqrt{1+3\cos^2\theta}} \\
\displaystyle \qquad{} = \frac{GM}{R_{\rm e}^2} \frac{\cos\theta}{\sqrt{1+3\cos^2\theta}} \left[ 
    3\omega^2 \sin^4\theta - \frac{2}{\sin^4\theta} \right],
\end{array}
\end{equation}
where we introduce a rotation frequency normalized by the Keplerian frequency at the outer boundary $\omega = \Omega / \Omega_{\rm K}(R_{\rm e})$, $R=R_{\rm e} \, \sin^2 \theta$ along the field line. 

The real funnel flow occupies a finite-size flux tube which we will characterize by its transverse surface area $A$ and polar-angle thickness $\delta$ calculated the same way as in \citetalias{AL23}. 
Density and velocity are considered constant on the surfaces orthogonal to the field lines. 
Stationarity and mass conservation imply that mass accretion rate is an invariant independent of $l$
\begin{equation}\label{E:Mass}
    \dot{M} = A \rho v = {\rm const}.
\end{equation}
Energy conservation including energy losses from the sides (but not diffusion along the field lines) may be written as (see \citetalias[section 2.3]{AL23}, but with the infall velocity $v>0$) 
\begin{equation}\label{E:energy}
    \frac{\diff }{\diff l} \left(\frac{4}{3}v u A + \left(\frac{v^2}{2}+\Phi  \right)\dot{M}\right) = \frac{2}{3\tau_\theta} \Pi c u + \left[\frac{4}{3} \frac{u}{\rho} +\Phi\right] \frac{\diff \dot{M}}{\diff l}\,.
\end{equation}
 Here  $\Pi$ is the radiating perimeter, 
\begin{equation}
    \tau_\theta = \varkappa \rho \delta
\end{equation}
is the transverse optical thickness,  $\varkappa \simeq 0.35{\rm cm^2\, g^{-1}}$ is the Thomson scattering cross-section,
$\delta \ll \Pi$ is the thickness of the flow in poloidal direction (see further details of the model geometry in AL23), and 
\begin{equation}
    \Phi = -\frac{GM}{R} + \frac{1}{2}\Omega^2 R^2\sin^2\theta = - \frac{GM}{R} \left( 1- \frac{\omega^2}{2}\sin^8\theta\right) 
\end{equation}
is the joint potential of the gravitational and centrifugal forces. 
We assume the equation of state is that of a radiation-pressure-dominated optically thick gas.

Taking into account equations (2--4) from \citetalias{AL23}, which are
\begin{equation}\label{E:delta}
        \delta = \frac{R_{\rm e} \sin^3{\thetamag}}{\sqrt{1+3\cos^2\thetamag}} \frac{\Delta R_{\rm e}}{R_{\rm e}}\,,
\end{equation}
\begin{equation}\label{E:A}
    A = 4\uppi a R_{\rm e } \Delta R_{\rm e} \frac{\sin^6\theta}{\sqrt{1+3\cos^2\theta}}\, ,
\end{equation}
\begin{equation}
    \Pi \simeq 8\uppi a R\sin\theta\, ,
\end{equation}
the geometrical factor in the right-hand side of Eq.~(\ref{E:energy}) may be written as
\begin{equation}\label{E:peridelta}
    \frac{\Pi}{\delta} \simeq \frac{8\uppi a R\sin\theta}{\delta} = \frac{8\uppi a \Re}{\Delta \Re} \sqrt{1+3\cos^2\theta}\, . 
\end{equation}
{Here $\Delta \Re = \delta(\Re)$ is the  width of the magnetospheric flow in the poloidal direction at its outer edge.
The azimuthal filling factor $a$ determines the width in the azimuthal direction, $2\uppi\, a\, \Re$.}

A transonic solution is expected to have a shock wave, whose position in the steady state depends on the heating and cooling balance in the flow below. 
In particular, the position of the shock was calculated in \citetalias{BS76}. 
At high enough mass accretion rates, its calculated position is above the sonic point, that raises the question of existence of a subsonic funnel flow extending without a shock all the way from the disc towards the surface of the star. 
Let us put aside the question about the azimuthal velocity in the disc and how does it change after crossing the Alfv\'{e}n surface and restore the structure of such a flow using Eqs~(\ref{E:Euler}), (\ref{E:Mass}), and (\ref{E:energy}), assuming that the whole flow is subsonic and neglecting the terms $\sim \Mach^2 = v^2/\cs^2$.

\subsection{Energy conservation}\label{sec:an:energy}

The results of the previous section allow us to write the energy equation for $\omega=0$  and $\dot M = \mathrm{const}$ as
\begin{equation}
\label{E:Energy_theta}
    \frac{\diff}{\diff \theta} \left[ \frac{4}{3} \frac{u}{\rho} - \frac{GM}{R}\right] = \frac{4}{3} \frac{a\Re}{\Delta \Re} \frac{\Re c^2}{GM} \frac{1}{\dot{m}} \left( 1+3\cos^2\theta\right) \,\sin \theta\, \frac{u}{\rho},
\end{equation}
where
\begin{equation}
\label{eq.mdot_def}
    \dot m = \frac{\dot{M}c^2}{L_{\rm Edd}} = \frac{\varkappa \dot{M} \, c}{4\uppi GM}\, .
\end{equation}
The first term on the left-hand side of Eq.~\eqref{E:Energy_theta} may be combined with its right-hand side. Taking advantage of the integral
\begin{equation}
    \int \left( 1+3\cos^2\theta\right) \sin \theta \diff \theta = - \cos\theta \left ( 1+\cos^2\theta\right) + \mbox{const},
\end{equation}
one obtains
\begin{equation}
    \displaystyle e^{-k \cos \theta \left( 1+ \cos^2\theta \right)} \frac{\diff}{\diff \theta} \left[ e^{k \cos \theta \left( 1+ \cos^2\theta \right)} \frac{u}{\rho} \right] = -\frac{3\cos\theta}{2\sin^3\theta} \frac{GM}{\Re},
\end{equation}
where we introduced dimensionless parameter
\begin{equation}\label{E:subsonic:k}
    k = \frac{a\Re}{\Delta \Re} \,\frac{\Re c^2}{GM} \frac{1}{\dot{m}} = \frac{a\Re}{\Delta \Re} \, \frac{3}{2}\frac{\Re}{R_{\rm sph}}\,,
\end{equation}
and 
\begin{equation}
    R_{\rm sph} = \frac{3}{8\uppi} \frac{\varkappa\dot{M}}{c}
\end{equation}
is the spherization radius \citep{SS73}. 
If the inner radius of the disc is smaller than $R_{\rm sph}$, the inner parts of the disc are poorly described by the standard thin-disc model. 
Small values of $k$ are thus associated with all the super-Eddington phenomena: large relative thickness of the accretion disc, mass loss in an optically thick radiation-driven wind, etc. 
Within the magnetospheric flow, small $k$ means larger impact of advection compared to radiation losses.

Let us express the solution in the form
\begin{equation}
\label{E:f_def}
    \frac{u}{\rho} = \frac{GM}{\Re} f(\theta),
\end{equation}
where normalized enthalpy $f(\theta)$ fulfils the equation
\begin{equation}\label{E:subsonic:f}
     {\rm e}^{-k \cos \theta \left( 1+ \cos^2\theta \right)} \frac{\diff}{\diff \theta} \left[ {\rm e}^{k \cos \theta \left( 1+ \cos^2\theta \right)} f(\theta) \right] = -\frac{3\cos\theta}{2\sin^3\theta}.
\end{equation}
The solution is more convenient to recover from the maximal polar angle $\theta_{\rm out} \simeq \uppi/2$ polewards (or inwards in terms of radial coordinate). 
If the value $f(\theta_{\rm out}) = f_{\rm out}$ is known, the formal solution is
\begin{equation}
\begin{split}
 \displaystyle &  f(\cos\theta) = \e^{-k \cos\theta (1+\cos^2\theta)} \,  \\ & \times \left[ f_{\rm out} \,  \e^{-k \cos\theta_{\rm out} (1+\cos^2\theta_{\rm out})} +  \frac{3}{2}\int_{\cos\theta_{\rm out}}^{\cos\theta} \e^{k x (1+x^2)} \frac{x\diff x}{\left( 1-x^2\right)^2}\right]\, .
 \label{E:f_theta_solution}
\end{split}
\end{equation}
The solution evaluated at the surface of the star allows us to calculate the advection parameter $\beta$ introduced by \citetalias{BS76} as
\begin{equation}
\label{E:beta-fout}
    \beta \equiv \frac{4}{3} \frac{R_*}{GM} \left( \frac{u}{\rho}\right)_0  = \frac{4}{3} \frac{R_*}{\Re} f(\cos\theta_0)\, ,
\end{equation}
where $\theta_0$ is the polar angle of the footpoint of the funnel flow. The value of $\theta_0$ follows from the dipolar shape of the field lines
\begin{equation}
\label{eq.sintheta0}
    \sin \theta_0 = \sqrt{\rstar/\Re}.
\end{equation}

We note that Eq.~(\ref{E:energy}) allows for the mass loss from the magnetic tube. 
If the specific energy of the matter being lost from the column is equal to the average one over the column cross-section, Eq.~(\ref{E:f_theta_solution}) is still valid, though $k$ is now a function of $\theta$ in the regions where mass is lost from the flow. 
Hereafter, we neglect the mass-loss term for simplicity, but a deeper study should take into account the structure of the column in the transverse direction and the details of the matter migration across the field lines. 

As it was shown in \citetalias{AL23}, at large mass accretion rates, when advection dominates the energy transfer within the column, the column is likely to lose mass at a finite altitude above the surface. 
Such \emph{vents} in accretion columns reduce the radiation pressure, but can also alter the observational appearance of the source. 
The matter lost from the columns will be accelerated to mildly relativistic velocities. 
This picture is confirmed by the numerical results of \citet{Takahashi-Ohsuga2017}, who simulate a rapidly accreting strongly magnetized NS launching outflows driven by radiation pressure.

 In \citetalias{BS76}, $\beta$ determines the X-ray luminosity of accretion columns (their equation 37). In the case of a vent present, advection parameter $\beta$ does not necessarily have the global physical meaning of the advected fraction of released gravitational energy. 
For large mass accretion rates ($\beta \sim 1$), most of the energy is probably lost from the flow as a result of mass leakage.

For any $k$ and $f_{\rm out}$, there is a unique solution $f(\theta)$, expressed by \eqref{E:f_theta_solution}. Parameter $k$ can be set following \eqref{E:subsonic:k}. However, to find $f_{\rm out}$ and  reconstruct the energy density profile, one needs to solve the momentum equation.

\subsection{Hydrostatic equilibrium}

The Euler equation  in the low-velocity limit is the equation of hydrostatic balance (Eq.~\eqref{E:Euler} without without the left-hand side):
\begin{equation}
    \frac{1}{\rho} \frac{\diff P}{\diff l} = \frac{GM}{\Re^2} \frac{\cos\theta}{\sqrt{1+3\cos^2\theta}} \left[ 3\omega^2 \sin^4\theta - \frac{2}{\sin^4\theta}\right]\, .
\end{equation}
Let us re-write this equation in terms of polar angle $\theta$, using Eq.~\eqref{E:dl}.
Substituting \eqref{E:f_def}, we obtain
\begin{equation}\label{E:subsonic:energy}
\displaystyle \frac{\diff}{\diff \theta} \ln u = \frac{3\sin \theta \cos\theta}{f(\cos\theta)} \left[ 3\omega^2 \sin^4\theta - \frac{2}{\sin^4\theta}\right].  
\end{equation}
For energy density $u$, we have the breakdown boundary condition at the NS surface \begin{equation}
    u = 3 \, u_{\rm mag}(\theta_0)\, .
    \label{E:inner_b_cond}
\end{equation}
At the outer edge of the magnetospheric flow, a similar condition should be satisfied, as the magnetospheric boundary is located at $\Re \sim R_{\rm A}$.  
We assume that  $u(\uppi/2) = u_{\rm mag}(\uppi/2)$ and obtain
 \begin{equation}
 \label{E:outer_b_cond}
     u(\uppi/2) = \frac{1}{3} \frac{u_{\rm mag}(\uppi/2)}{u_{\rm mag}(\theta_0)} u(\theta_0) \simeq \frac{1}{3} \frac{\sin^{12}\theta_0}{1+3\cos^2\theta_0} u(\theta_0)\, .
 \end{equation}
The factor on the right-hand side reflects the change in the magnetic field density along the field line. 
 
Equation~\eqref{E:subsonic:energy} does not have an analytical solution, but may be integrated numerically, taking into account restrictions from the physics of the problem,  described in the following subsections. 
The only free parameter of the problem, $f_{\rm out}$, is found by the shooting method minimizing the residual in the energy density in~Eq.~(\ref{E:subsonic:energy}).

\subsection{Limits on  pressure and  enthalpy}\label{sec:an:lim}

Requiring $P_{\rm rad}\leq P_{\rm mag}$, or $u\leq 3u_{\rm mag}$, everywhere in the flow, from \eqref{E:subsonic:energy} we have
\begin{equation}
\label{E:pressure_limit}
    \frac{\diff }{\diff \theta} \ln u_{\rm mag} > \frac{3\sin \theta \cos\theta}{f(\cos\theta)} \left[ 3\omega^2 \sin^4\theta - \frac{2}{\sin^4\theta}\right]. 
\end{equation}
Approximating the radial dependence of the magnetic energy density as $u_{\rm mag} \propto (1+3\cos^2\theta ) R^{-6} \propto (1+3\cos^2\theta)\sin^{-12}\theta$ and assuming $\omega =0$, we get a condition for 
force-free approximation validity within the flow 
\begin{equation}
\label{E:f_limit}
    f < \frac{1}{\sin^2\theta} \frac{1+3\cos^2\theta}{3+5\cos^2\theta} \simeq  \frac{\Re}{2R}\,.
\end{equation}
This is simply a requirement that the sound speed  in the flow is everywhere about- or sub-Keplerian. 
{The restriction seems natural}, though the flow does not necessarily need to be gravitationally bound (Bondi flow being a good example). 
Violating this restriction would also mean that the radial slope of energy density is shallower than that of the magnetic field. 
Applied to the NS surface, condition \eqref{E:f_limit} coincides with the condition for the vent opening just at the surface of NS, $\beta < 2/3$, found in \citetalias{AL23}. 
In the opposite case, a vent opens above the surface of the NS.

The semi-analytic solution does not have to obey the pressure limit \eqref{E:pressure_limit}.
As we will see in the following subsection, requirement~\eqref{E:f_limit} is practically never fulfilled for $k \lesssim 1$, meaning that all the solutions should have vents above the NS surface. 
In the setup of \citetalias{AL23}, the mass-flow rate is controlled by the breakdown condition along the tube. 
Given all the considerations above, one would expect a steady-state flow to have a normalized enthalpy dependence on radius to be determined by the energy balance equation \eqref{E:f_theta_solution}, and the energy density to be $u\sim u_{\rm mag}$ at all the radii.

\subsection{Fully advective super-Eddington regime}\label{sec:subsonic:smallk}

Small $k \ll 1$ is a reasonable approximation for a super-Eddington regime when the spherization radius is larger than $R_{\rm e}$, see Eq.~\eqref{E:subsonic:k}. 
In zeroth order in $k$, Eq.~(\ref{E:subsonic:f}) may be rewritten as
\begin{equation}
\frac{\diff f}{\diff \theta} = - \frac{3\cos\theta}{2\sin^3\theta}, 
\end{equation}
which is easily integrable as
\begin{equation}\label{E:lowk:f}
    f(\theta) \simeq \frac{3}{4} \frac{\Re}{R_*} \beta + \frac{3}{4}\left( \cot^2\theta - \cot^2\theta_0\right).
\end{equation}
Alternatively, one can write down $f$ as a function of radius as
\begin{equation}\label{E:f:R}
    f(R) \simeq \frac{3}{4} \frac{\Re}{\rstar} \left( \frac{\rstar}{R} - 1 +\beta\right).
\end{equation}
In physical units, this corresponds to 
\begin{equation}\label{E:f:phys}
    \frac{u}{\rho} \simeq \frac{3}{4} \frac{GM}{R}.
\end{equation}
To keep enthalpy $f$ positive everywhere, one must ensure that 
\begin{equation}
\label{E:beta_req}
\beta \gtrsim 1- \rstar/\Re,
\end{equation}
which certainly restricts the size of the magnetosphere. 
It also allows for a very narrow range of $\beta \sim 1$, which does not necessarily imply that the flow is fully advective, given that some mass and energy are lost somewhere above the surface.
Assuming $\beta = 1$, we arrive at the limiting outer boundary condition  $f_{\rm out} = 0.75$ and the specific solution
\begin{equation}\label{eq.fk0}
    f_{k \to 0} \simeq \frac{3}{4} \frac{\Re}{R}\, .
\end{equation}
Substituting $f_{k \to 0}$ into Eq.~(\ref{E:subsonic:energy}) yields, for $\omega=0$,
\begin{equation}\label{E:u}
    u \simeq \left( 1 + \frac{R_*}{\Re} \left( \cot^2\theta - \cot^2\theta_0\right)\right)^{4} u(\theta_0)\, .
\end{equation}
This scaling may be obtained and understood also in terms of the equation of state. 
Setting $k=0$ means neglecting radiation losses, hence in the low-$k$ regime $f \propto u/\rho \propto u^{1/4}$.
As a consequence, radiation pressure in this solution exceeds magnetic pressure practically everywhere: solution \eqref{eq.fk0} violates condition \eqref{E:f_limit}.
As we suggested in \citetalias{AL23}, this violation may be resolved by a vent formation (see also discussion in Sect.~\ref{sec:an:energy} after Eq.~\ref{eq.sintheta0}). 
This mass loss leaves intact solution for $f$, while the   
values of the energy density become lower than predicted by Eq.~(\ref{E:u}) and close to $u_{\rm mag}$.

\subsubsection{Physical properties of the advective solution}\label{sec:subsonic:smallk:vel}

If we parametrize,  
\begin{equation}
    u = \frac{B^2(R)}{8\uppi} \psi(\theta)\, ,
\end{equation}
where $\psi(\theta)\sim 1$ is a weak function of the angle, it is possible (using Eq.~\ref{E:f:phys} and mass conservation) to calculate the density and velocity profiles as
\begin{equation}\label{E:obs:rho}
    \rho = \frac{4\psi}{3} \frac{R}{GM} u_{\rm mag}\, ,
\end{equation}
\begin{equation}\label{E:obs:v}
    v = \frac{\dot{M}}{A \rho} = \frac{3}{4\psi} \frac{GM\dot{M}}{R} \frac{1}{A\, u_{\rm mag}}\, .
\end{equation}
Radial dependences for the two quantities are $\rho \propto R^{-5}$ and $v\propto R^{2}$. {Here, we have assumed that the mass inflow rate is constant, $\dot{M}=$\,const. This assumption is going to be violated inside and below the vent, leading to sinking velocities smaller than predicted.}

The condition of the outer velocity being sub-Keplerian, or, more precisely, less than the free-fall velocity $\sqrt{2}\,v_{\rm K}$, corresponds to 
\begin{equation}
    \left( \frac{v}{v_{\rm K}}\right)_{R = \Re} \simeq \frac{3}{4\,\psi} \, \frac{\Re}{a\Delta \Re} \, \ximag^{7/2} < \sqrt{2},
\end{equation}
where $\ximag = \Re / R_{\rm A}$.
The condition above effectively sets a lower limit for the transverse size of the flow: $a\Delta \Re/\Re \gtrsim 0.05$, if one adopts $\ximag = 0.5$.

For all the calculations here, we have assumed that the flow is optically thick. 
{The transverse optical depth of the flow}
\begin{equation}
    \tau \simeq \varkappa \rho \delta\, ,
\end{equation}
where $\delta$ is given by Eq.~\eqref{E:delta}. Taking into \eqref{E:obs:rho},
\begin{equation}
    \tau = \frac{8\sqrt{2}\psi}{9} \ximag^{-8} \frac{R_{\rm sph}}{\sqrt{R_{\rm g} R_{\rm A}}} \frac{\sqrt{1+3\cos^2\theta}}{\sin^7\theta} 
\end{equation}
with 
\begin{equation}
    R_{\rm g} = \frac{GM}{c^2}\, .
\end{equation}
The overwhelmingly strong dependence of the optical depth on $\ximag$ means that the former may be consistently high even for reasonably low mass accretion rates. 
Assuming $\ximag = 0.5$ and $\Delta \Re / \Re \sim 1$, we expect $\tau \gg 1$ for the entire range of radii if
\begin{equation}\label{E:Rsph:gtr}
    R_{\rm sph} \gtrsim \sqrt{R_{\rm g} R_{\rm A}}\, .
\end{equation}
If $R_{\rm sph} > \Rm \sim R_{\rm A}$, one would expect the Eddington limit to be violated in the inner parts of the disc, the latter becoming geometrically thick and possibly launching optically thick winds (see \citealt{CLAP}). 
The limit for $\dot{M}$ set by Eq.~(\ref{E:Rsph:gtr}) is lower, meaning there is also a range of mass accretion rates at which an optically thick subsonic flow exists inside a sub-Eddington disc. 

\begin{figure*}[h!]
\includegraphics[width=1.0\textwidth]{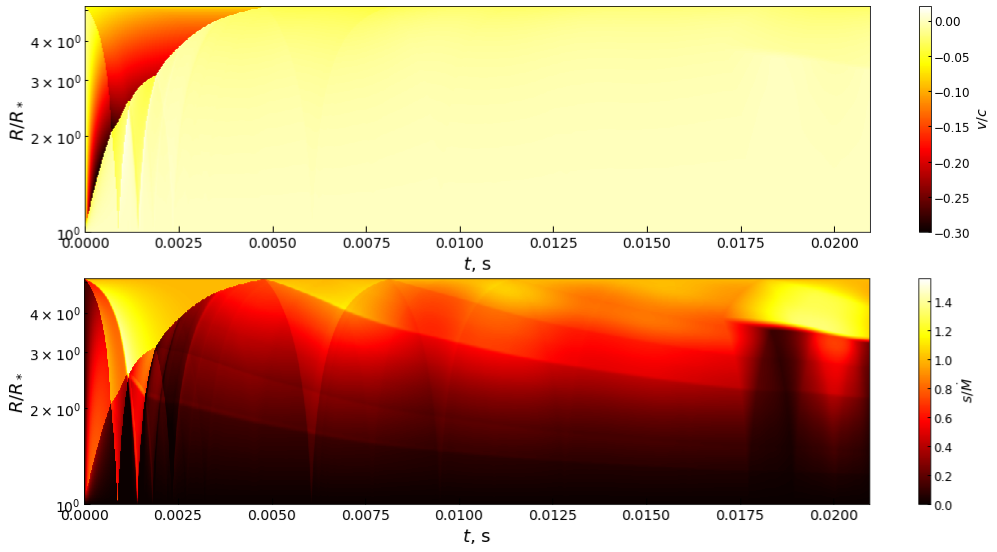}
 \caption{ Local velocity and mass accretion rate $s = \rho v A$ plotted as functions of radius and time for the simulation \fidu. Only the initial $\sim 0.02$~s of the evolution are shown. 
 }\label{fig:q2}
\end{figure*}

\section{Numerical simulations}\label{sec:sim}

To check the semi-analytic solution, we ran one-dimensional time-dependent MHD simulations. 
We used the code {\tt HACol}\footnote{The code is freely available at \url{https://github.com/pabolmasov/HACol.git}.}, described in \citetalias{AL23}, with a few updates. 
In particular, pressure at the outer boundary was set to $u_{\rm mag}/3$, allowing comparison to the analytic solution. 
In \citetalias{AL23}, we  assumed equality between the energy densities which gives $P$ between $u_{\rm mag}/3$ and $2u_{\rm mag}/3$. 
{The difference is in the contribution of the gas pressure, and never exceeds one per cent in the simulations presented here but may be much larger in a transonic flow above the shock.}  

We have calculated two models that differ only in the condition for mass loss. \glaccept{In the fiducial model \fidu, we follow the original condition for mass, momentum, and energy losses \citepalias[section 2.3]{AL23}. In model \zeroloss, we turn off the mass losses and the associated loss terms for energy and momentum, thus allowing the flow to accumulate indefinitely large amounts of mass and heat.} 
In both simulations, $\dot m = 300$, $\mu = 10^{29}\Gcmc$, $a=1/4$, $\Delta \Re/\Re=1/4$, which implies the size of the magnetosphere $\Re \simeq 5.3\rstar$, $\theta_0 \simeq 26^\circ$, and $k \simeq 0.085$.
The advection parameter of the corresponding BS solution is $\betaBS \approx 0.92$. 

The simulations were run for $0.5$\,s, with is of the order of several tens of the column replenishment times ($t_{\rm r} =2500~{\rm s}\, B_{12}^2 \epsBS \delta_*/\rstar \sim 0.01$~s), see \citetalias{AL23} and \citetalias{variable-opacity}. \footnote{In our units, $\epsBS =2 \,a/\dot m \,(\rstar/R_{\rm g}) \sqrt{\rstar/\Re}$.}

Fig.~\ref{fig:q2} presents time-radius distribution for the velocity and the normalized accretion rate in model \fidu. 
We show only the initial 40 milliseconds, as the subsequent evolution is much more stable\footnote{More figures may be found at \url{https://github.com/HACol-data/shock-free.git}.}. 
During the first milliseconds, it is possible to see moving shocks and nearly free-falling matter (Keplerian velocity varies from roughly $0.06$c on the outer edge to $0.2$c near the surface). Then, the whole column is involved in sub-Keplerian and sub-sonic sinking at a speed of $v\lesssim 0.02$c. 
The leakage of mass form the columns occurs in a wide interval of heights.  
A vent forms at $t\simeq 0.02$s and does not fully stabilize: up to $t\simeq 0.5$s, the  column is involved in vertical oscillations with at least two different frequencies.
We will discuss the properties of these pulsations in Sect.~\ref{sec:obs:pulse}.

\begin{figure}[]
\includegraphics[width=0.44\textwidth]{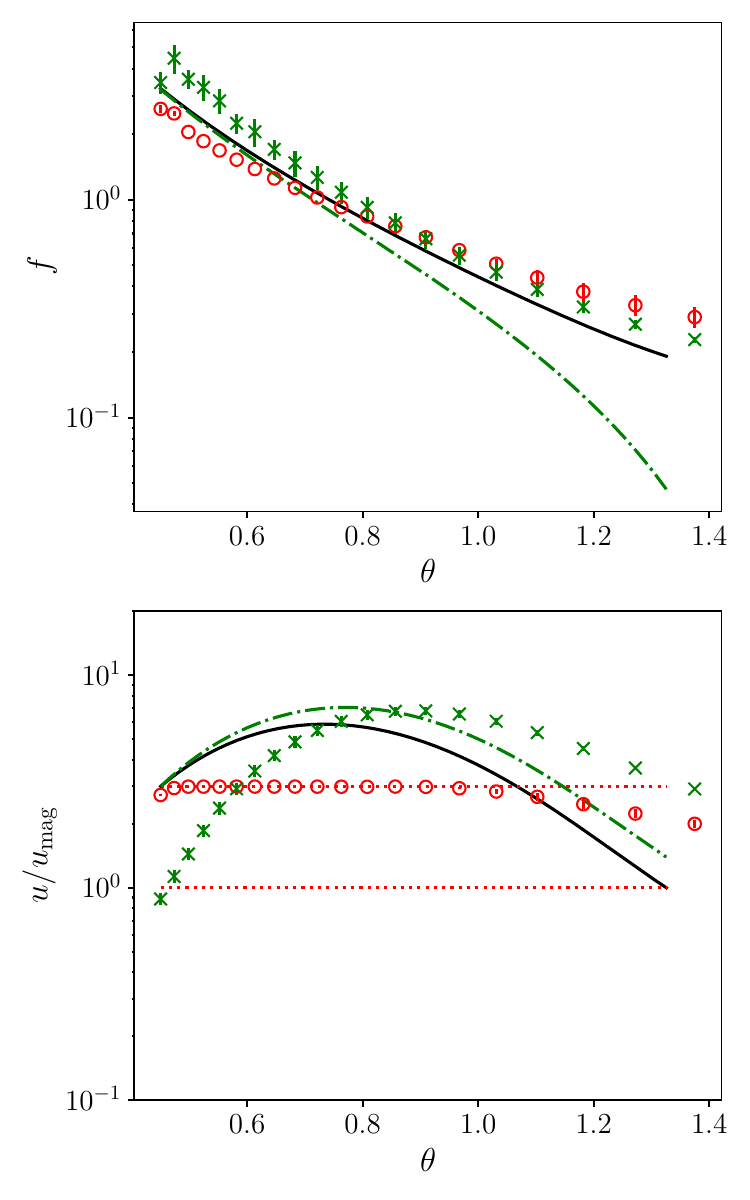}
 \caption{ Normalized enthalpy $f$ (upper panel) and energy density $u$ in the magnetic energy density units (lower panel). Semi-analytic subsonic solution produced by numerical integration of Eqs.~(\ref{E:subsonic:f}) and \eqref{E:f_theta_solution} is shown by the solid black lines. Open red circles show the numerical simulation \fidu~($k\simeq 0.085$), the simulation \zeroloss is shown with green crosses. Green dot-dashed lines are the low-$k$ approximation solutions from Sect.~\ref{sec:subsonic:smallk}. The eigenvalues of the solution are $f_{\rm out} =  0.19$ and $\beta = 0.82$. 
 }
\label{fig:uintT}
\end{figure}
In Fig.~\ref{fig:uintT}, we compare the two stationary analytic solutions (the one obtained by numerical integration of Eqs.~(\ref{E:subsonic:f}) and (\ref{E:subsonic:energy}) and the low-$k$ analytic approximation) with  the late-time snapshots of both time-dependent simulations, \fidu and \zeroloss.
Simulation \fidu\ reproduces well the overall shape of $f(\theta)$ (upper panel), but does not follow the peak of the energy density distribution (lower panel), as the code restricts the maximal energy density (see Sect.~\ref{sec:an:lim} for details).  
Thus, as it was suggested in Sect.~\ref{sec:subsonic:smallk}, mass loss does not affect appreciably the solution for normalized enthalpy $f(\theta)$.

\section{Observational implications and discussion}\label{sec:obs}

\subsection{Luminosity and spectrum}
\label{sec.luminosity}
The local radiation flux through the wall of the column is
\begin{equation}\label{E:obs:F}
    F \simeq \frac{cu}{3\tau} \simeq \frac{c}{4\varkappa \delta} \frac{GM}{R}\, .
\end{equation}
The integral luminosity of the columns is obtained by  integrating the flux over the surface
\begin{equation}
    L = \int F \Pi \diff l = \frac{GMc}{4\varkappa} \int \frac{\Pi}{\delta} \frac{\diff l}{R}.
\end{equation}
Substituting the expressions for $\Pi/\delta$ (Eq.~\ref{E:peridelta}) and $\diff l$ (Eq.~\ref{E:dl}), one arrives at
\begin{equation}
L = \frac{1}{2} \frac{a \Re}{\Delta \Re} L_{\rm Edd} \left[ 2\ln \frac{1+\cos\theta_0}{1-\cos\theta_0} - 3 \cos\theta_0\right],   
\end{equation}
where $\theta_0$ is given by \eqref{eq.sintheta0}.
Large magnetospheres may shine brighter than the Eddington limit by a factor of 
\begin{equation}
    \frac{L}{L_{\rm Edd}} \simeq \frac{a \Re}{\Delta \Re} \ln \left( \frac{4}{e^{3/2}} \frac{\Re}{\rstar}\right)
\end{equation}
Or, the maximum bolometric luminosity of the columns in   the advective shock-free  regime is 
\begin{equation}
\label{eq.Lcol_max}
L \simeq \frac{a \Re}{\Delta \Re} L_{\rm Edd} \left[ 4.5+ 0.29\, \ln\left(\frac{\mu_{30}^2}{\dot m} \right)\right].
\end{equation}
Here $\dot m$ is defined according to \eqref{eq.mdot_def} and $\mu_{30} = \mu/10^{30}$~G~cm$^3$. 


The local temperature may be calculated using Eq.~(\ref{E:obs:F}) as
\begin{equation}
\begin{array}{l}
  \displaystyle  T_{\rm eff} \simeq 
  \left( \frac{GMc}{4\sigma_{\rm SB} c\Re^2} \frac{\Re}{\Delta \Re}\right)^{1/4} \left( 1+3\cos^2\theta\right)^{1/8} \sin^{-5/4}\theta \\
\displaystyle \qquad{}    \sim 0.3 {\rm keV} \left( \frac{\Re}{10^8\rm cm} \right)^{-1/2} \sin^{-5/4}\theta \  \propto R^{-5/8} \, .\\
\end{array}
\end{equation}
The spectrum may be calculated as a multi-coloured black body with a temperature varying by about a factor of $\sin^{-5/4}\theta_0 = (\Re/\rstar)^{5/8}$. 
The geometry of the radiating region is complicated, that likely affects its observed spectral variability. 
For an observer located close to the magnetic axis, the observed spectral shape in the multi-colour approximation is analogous to the spectrum of an accretion disc with a  generalized power-law temperature distribution, a $p$-free disc ( $T_{\rm eff} \propto \varpi^{-p}$, see \citealt{2000PASJ...52..133W}) with $p=5/12$. 
The resulting spectrum is soft, $\nu L_\nu \propto E^{-4/5}$ for $T_{\rm out} \lesssim E \lesssim T_{\rm in}$, where $T_{\rm out} = T_{\rm eff}(\theta = \uppi/2)\sim 0.3\keV$ and $T_{\rm in} = T_{\rm eff}(\theta = \theta_0) \sim 5\keV$ for $\Re = 10^8\cm$  and $\theta_0 \simeq 0.1$.

For the observational properties of the source, it is also important that the observed radiation from the column is likely to be scattered both in situ (because scattering cross-sections dominate over true absorption) and non-locally, especially if the azimuthal coverage of the flow is high $a\sim 1$. 
As the result, the observed radiation is going to be Comptonized and mildly geometrically channelled (part of the radiation is concentrated in a solid angle $\sim (1-\cos\theta_0)$), which leads to spectrum hardening and larger isotropic luminosities. We further discuss the consequences of non-local scattering later in Sect.~\ref{sec:disc:beaming}.

\begin{figure}
\includegraphics[width=\linewidth,trim={0cm 0cm 0cm 0cm},clip]{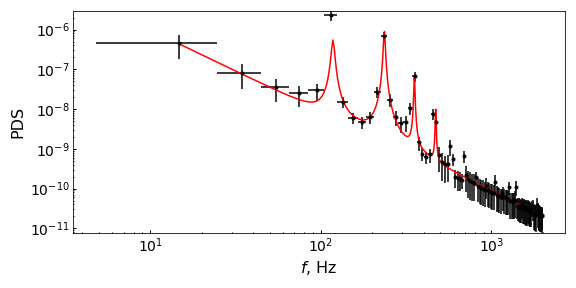}
\caption{Integral PDS of the model \fidu\ in the time range $t = 0.4-0.9$~s (about 50000 evenly spaced data points). The red curve shows a fit with a power-law and four Lorentzian components with free amplitudes and widths. The first Lorentzian central frequency is $f_0 = 118\pm 0.4$Hz, the others have frequencies locked to its multiples ($2f_0$, $3f_0$, and $4f_0$, higher harmonics are visible but not included in the fit). 
 }
 \label{fig:PDS}
\end{figure}

\subsection{Variability modes}\label{sec:obs:pulse}

The sound travel time between the outer and inner edges is
\begin{equation}
    T_{\rm s} = \int_{\theta_0}^{\uppi/2} \frac{\diff l}{\cs}, 
\end{equation}
where $\diff l$ may be expanded using Eq.~(\ref{E:dl}). 
The speed of sound may be expressed using the solution for $f$. 
In the low-$k$ (adiabatic) limit, $f(\theta)$ is approximated by Eq.~(\ref{eq.fk0}), which results in
\begin{equation}
\begin{array}{l}
 \displaystyle   T_{\rm s} = \frac{\Re^{3/2}}{2\sqrt{GM}} \int_{\theta_0}^{\uppi/2} \frac{\sqrt{1+3\cos^2\theta} \sin \theta \diff \theta}{\sqrt{f}} \\
\displaystyle \qquad{}  \simeq \frac{\Re^{3/2}}{2\sqrt{GM}} \int_{\theta_0}^{\uppi/2} \sqrt{1+3\cos^2\theta} \sin^2 \theta \diff \theta.
\end{array}
\end{equation}
The integral can be expressed as \citep[2.583/4]{gradshteyn2007}
\begin{equation}
\begin{array}{l}
    \displaystyle T_{\rm s} \simeq \frac{\Re^{3/2}}{9\sqrt{GM}} \left[3\sin\theta_0\cos\theta_0 \sqrt{1-\frac{3}{4}\sin^2\theta_0} \right. \\
   \displaystyle \left. + \EllipticF{\uppi/2}{\frac{3}{4}}-\EllipticF{\theta_0}{\frac{3}{4}} + 2\EllipticE{\uppi/2}{\frac{3}{4}}-2\EllipticE{\theta_0}{\frac{3}{4}}\right],\\
    \end{array}
\end{equation}
where $\EllipticF{x}{m}$ and $\EllipticE{x}{m}$ are elliptic integrals of the first and the second kind, respectively. 
Approximately, to the zeroth order in $\theta_0$ and after rounding off the error to an accuracy of about 2 per cent, 
\begin{equation}\label{E:time:sound}
    T_{\rm s} \simeq 0.51 \frac{\Re^{3/2}}{\sqrt{GM}} \simeq 0.081 T_{\rm K},
\end{equation}
where $T_{\rm K}$ is Keplerian rotation period at $\Re$. If the disc is close to co-rotation, $T_{\rm K}$ is close to the spin period of the NS.
In the PDS, this sonic mode will produce a peak at a frequency about ten spin frequencies.

In the model \fidu,  the measured power-density spectrum (PDS, see Fig.~\ref{fig:PDS}) has a peak at a frequency lower by about an order of magnitude, $f = 100\pm20$Hz, and several peaks corresponding to its first harmonics.  For the parameters used in our numerical simulations, $f_{\rm s} = 1/T_{\rm s} \simeq 2{\rm kHz}$ from \eqref{E:time:sound}.
While it is difficult to detect the sonic peak in the PDS, it is probable that sonic modes are present (the waves themselves are seen in the lower panel of Fig.~\ref{fig:q2}), but coexist with a relaxation cycle of mass accumulation and loss that dominates the overall variability.

 The relevant time scale to explain oscillations in Fig.~\ref{fig:PDS} could  be the replenishment time $t_{\rm r} = M_{\rm col}/\dot{M}$ introduced by \citetalias{AL23}. 
$M_{\rm col}$ here is the mass of the column. 
The physical meaning of this time scale is the time needed to replace all the matter the column is composed of. 
While for the shock-mediated case, this time is usually much longer than the dynamical time of the column $T_{\rm K}$, they become comparable for the subsonic solution. 
Using the approximations of Sect.~\ref{sec:subsonic:smallk} (and setting $\psi = 1$), one can estimate the mass as
\begin{equation}
\begin{array}{l}
\displaystyle    M_{\rm col} = \int A \rho \diff l \simeq \frac{2\uppi }{3} \,a\,\frac{\Delta \Re}{\Re} \frac{\mu^2}{GM \Re^2} \int_{0}^{\cos\theta_0} \frac{1+3x^2}{(1-x^2)^2} \, \diff x \\
  \displaystyle  \qquad{} \simeq \frac{4}{3} \,a\, \frac{\Delta \Re}{\Re} \frac{\mu^2}{GM \Re \rstar} \, .
\end{array}
\end{equation}
For the replenishment time, this implies
\begin{equation}\label{eq.t_r}
    t_{\rm r} \simeq \frac{64}{3\uppi} a\frac{\Delta \Re}{\Re}  \frac{\Re}{\rstar} T_{\rm K}(\Re). 
\end{equation}
A narrow funnel flow in a small magnetosphere is likely to have a relatively short replenishment time, of the same order as the sound-propagation time given by Eq~(\ref{E:time:sound}). For our set of parameters, replenishment time corresponds to the frequency of $f_{\rm r} \simeq 80$Hz, reasonably consistent with the observed oscillation cycle.

\begin{figure*}
\centering
\adjincludegraphics[width=1.0\textwidth,trim={1.5cm 0.5cm 0.6cm 0cm},clip]{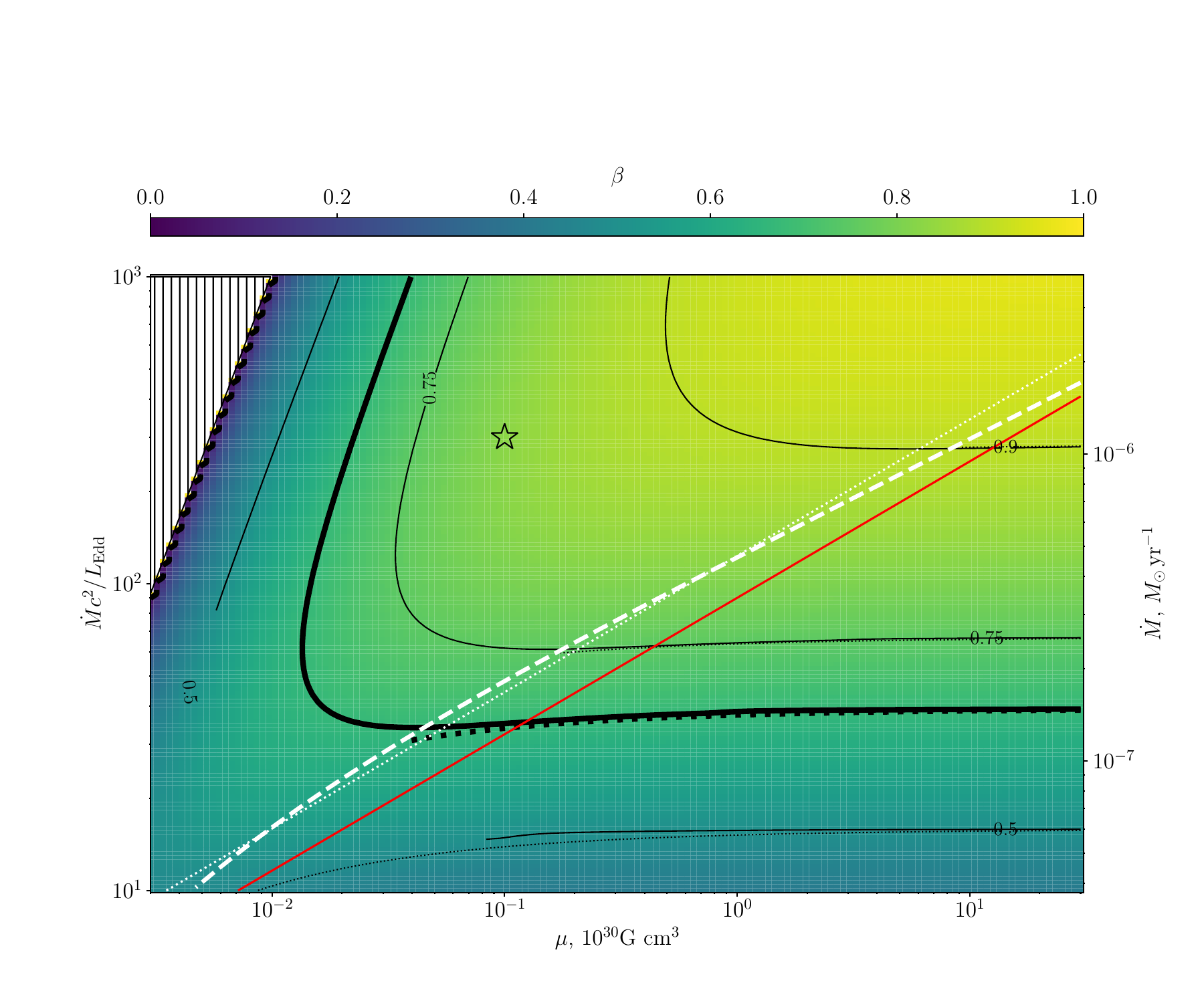}
 \caption{
  Estimated advection efficiency $\beta$ (see Eq.~\ref{E:beta-fout}) for the subsonic flow as a function of dimensionless mass accretion rate $\dot{m}$ and magnetic moment $\mu$.   
 Solid black contour lines are isolines of $\beta$ for the subsonic solution. 
 Black dotted lines are $\betaBS$ calculated for the shock-mediated solution of \citetalias{BS76}. 
 The thick black lines are the contour lines $\beta = 2/3$ for the two solutions (dotted is for the case with a shock).
 The white dashed line corresponds to the shock position $R_{\rm shock} = \Re$ in the shock-mediated analytical solution. 
 Below this line, the solution with a shock is possible. 
 Below the white dotted line, the low-$k$ solution is inapplicable, as condition \eqref{E:beta_req} is not fulfilled.
 Black vertical hatches mark the region without a magnetosphere ($\Re < \rstar$).
 Parameters of the numerical solution described in Sect.~\ref{sec:sim} are shown by a star. 
Along the red solid line, the radius of the magnetosphere equals spherization radius ($R_{\rm sph} = \Re$). For this plot, we assume $a=1/4$, $\Delta \Re / \Re = 1/4$. 
 }
 \label{fig:blimits}
\end{figure*}

\subsection{The population of subsonic sources}\label{S:population}

Depending on the mass accretion rate, properties of the magnetic field, and the geometrical parameters of the system, a high-rate magnetospheric accretion flow may be closer to the classical shock-mediated solution or to the subsonic solution found in this paper. 
The applicability limits for each of them involve a number of factors so far poorly constrained, including the electron-scattering opacity variations in magnetic field. As shown in \citetalias{variable-opacity}, the scattering opacity within the column depth is expected to remain at or above the classical Thomson value in the shock-free regime.
Assuming the Thomson opacity is applicable,
we summarize the restrictions and requirements for the shock-mediated and subsonic solutions in Fig.~\ref{fig:blimits}.
The colour-coded values of $\beta$ are calculated using the value of the normalized enthalpy at the inner edge of the magnetospheric flow in the  semi-analytic solution, see Eq.~\eqref{E:beta-fout}.
We also show some specific isolines of $\beta$ calculated this way for the subsonic solution (solid lines) and for the shock-mediated solution of \citetalias{BS76} (dotted). 
Within the applicability limits of the latter, the isolines show excellent consistency, suggesting that the value of $\beta$ is not sensitive to the presence of the shock but instead is defined by the inner boundary conditions and the flow just above the stellar surface.

The vertical black hatches in the upper left corner show the region where NS does not have a magnetosphere. 
The white dotted line marks the upper boundary of the region  where low-$k$ approximation becomes invalid because $\beta \leq 1-\rstar/\Re$, see Eq.~\eqref{E:beta_req}. 
Radiation losses in this region of the diagram are important. 
Above the white dotted line in Fig.~\ref{fig:blimits}, the subsonic steady-state solution  is possible and is highly advective. 
Below the white thick dashed line, the shock-mediated steady-state solution is possible (the expected location of the shock is within the magnetosphere). 
The fact that this line is close to the consistency limit of the subsonic solution means that different regions on the $\mdot - \mu$ plane are clearly associated with different accretion regimes.
For example, an object with $\mu = 10^{29}\Gcmc$ is likely to have the  advective accretion column in the mass accretion rate range $\sim 10^{-8}- 10^{-6}\Msunyr$ (luminosities up to $\sim 6\times 10^{39}\ergl$) and becomes a subsonic accretor at a larger accretion rates. 
One cannot exclude a scenario when a fixed set of global parameters allows for a number of different steady-state solutions or does not have a steady solution at all.

{There are observational indications \citep{2019ApJ...871..231H} that the donor stars in ULX pulsars are moderately massive \hbox{($\sim 10\,\Msun$)}. 
 Population synthesis \citep{Kayanikhoo+2025} predicts a population of NS X-ray binaries with intermediate-mass ($1-10\,\Msun$) slightly evolved donor stars and mass transfer on the thermal time scale of the donor. 
The life time of such a source exceeds the Hall decay time  of an initially magnetar-scale magnetic field (about $10^4$ years, see for instance \citealt{2007A&A...470..303P}), implying that the majority of the sources in this population are either born above the white lines in Fig.~\ref{fig:blimits} or migrate there relatively early.

\begin{figure}
\centering
\adjincludegraphics[width=1.0\columnwidth,trim={0cm 0cm 0cm 0cm},clip]{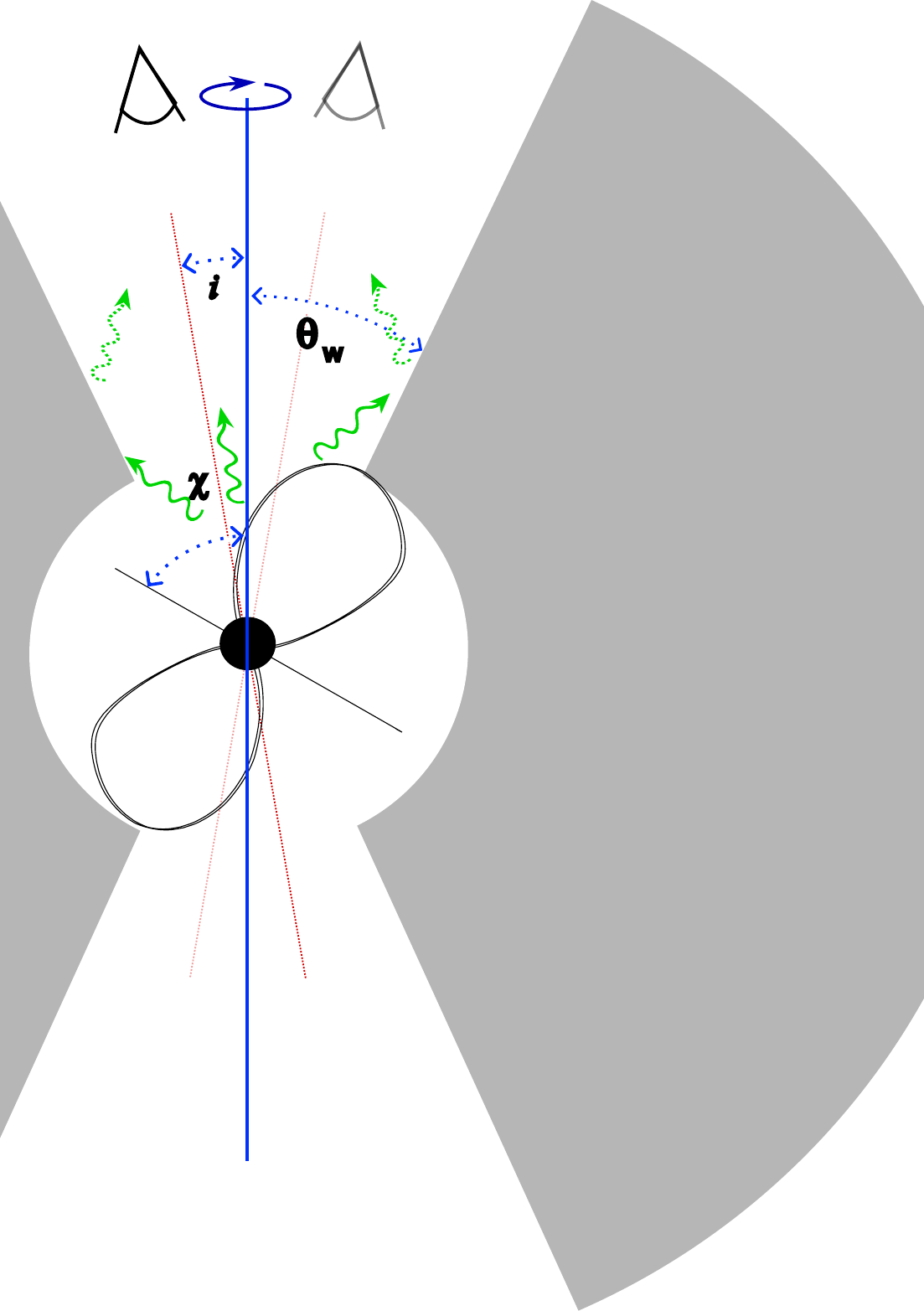}
 \caption{
Nested beaming. 
Magnetospheric flow (shown schematically with a double solid black line) radiates emission with a broad beam pattern, which is seen by the observer at different spin phases at different angles, from $\chi-i$ to $\chi+i$. Besides, some of the radiation is scattered by the wind (shown with the grey filling). Photons radiated by the magnetospheric flow are schematically shown with solid green wiggly lines, secondary photons scattered by the wind with dotted ones. Straight dotted red lines show the direction towards the observer at two different spin phases. Dotted blue arches mark the characteristic angles of the problem: magnetic obliquity $\chi$, inclination $i$, and wind half-opening angle $\theta_{\rm w}$. }
 \label{fig:nested_sketch}
\end{figure}
Many rapidly accreting NSs are thus expected to accrete in the subsonic regime, with a small minority of strongly magnetised objects having relatively large magnetospheres and accretion shocks.

The indication for strong, of the order $\sim 10^{13}-10^{14}$~G,  magnetic fields in the pulsating ULXs studied so far~\citep{2016MNRAS.457.1101T,2020ApJ...891...44B,2023A&A...672A.140F}   hints that pulsations are detectable primarily in the strongly magnetised regime, where the magnetosphere remains sufficiently extended to sustain an accretion shock. {On the other hand, strong magnetic fields favour  short columns as discussed in  \citetalias{variable-opacity}.}


\subsection{Nested beaming and pulse fractions}\label{sec:disc:beaming}

As one can see in Fig.~\ref{fig:blimits}, violation of the Eddington limit in the disc practically coincides with the subsonic regime in the magnetosphere.
For super-Eddington accretion in the disc, a conical non-relativistic outflow from the disc is expected~\citep{SS73, 1999AstL...25..508L, 2007MNRAS.377.1187P, CLAP}, which affects the observational properties of subsonic XRPs in multiple ways.
The analysis of this sub-section is applicable to any pulsating NS surrounded by a super-Eddington accretion disc.

The magnetosphere itself has a beam pattern of a complicated shape, which is determined by a number of poorly constrained physical and geometrical effects. 
For simplicity, let us assume that the beam is axisymmetric with respect to the magnetic axis, but the axis is misaligned with the rotational axis by angle $\muang$.
Detectability of the pulsations depends on the pulsation amplitude, commonly expressed in terms of a pulse fraction 
\begin{equation}
    PF = \frac{F_{\rm max} - F_{\rm min}}{F_{\rm max}+F_{\rm min}}\, ,
\end{equation}
where $F_{\rm min, \, max}$ are  the maximal and minimal fluxes, respectively,  measured   within a spin period.

The wind from the disc affects the effective beam shape and other observed properties of the object in multiple ways. 
First, the object is detected directly by the observer only for inclinations $\incl \lesssim \conus$, where $\conus$ is the half-opening angle of the wind. 
The opening angle of the wind may be as small as $\conus \sim 15\deg$, as it was estimated for the Galactic super-Eddington source \hbox{\object{Cyg X-3} }~\citep{2024NatAs...8.1031V}.
As a result, the source is observed in a narrow range of magnetic inclinations (from $\muang - \incl$ to $\muang + \incl$, see Fig.~\ref{fig:nested_sketch}), which suggests weak variations of the flux. 

The ionized wind works as an efficient reflector, intercepting and scattering a large fraction of the radiation coming from the central source.  
The contribution of the scattered emission depends weakly on the spin phase. 
Assuming that the scattered emission flux $F_{\rm sc}$ is independent of the spin phase, an estimate can be made for the decreased pulse fraction
\begin{equation}
    PF_{\rm sc} = \frac{PF}{1 + 2 F_{\rm sc} / \left( F_{\rm min} + F_{\rm max}\right)}.
\end{equation}
If $F_{\rm sc} \sim (F_{\rm min}+F_{\rm max})/2$, then the pulse fraction is reduced roughly by a factor of $2$.
Conventional approach to geometrical beaming \citep{king01,AKK} suggests that scatterings efficiently focus all the scattered radiation within the funnel.
Hence, a more accurate estimate of $F_{\rm sc}$ should include the geometrical collimation factor $b \sim 1/(1-\cos \conus)$, which is about a factor of $30$  for $\conus = 15\deg$. 
On the other hand, numerical Monte-Carlo simulations performed by \citet{2026arXiv260529487A} show that the emission scattered by the funnel walls has a much broader beam pattern.
As a result, a large fraction of observers are going to detect a source which is still X-ray bright (and classified as ultra-luminous) but has a strongly suppressed pulsed fraction, as most of the radiation is scattered by the funnel walls at least once. 

A strong observational prediction of this explanation for the non-pulsating ULX population is that the emission in such objects should be additionally polarized by scattering in the optically thick wind of the disc. 
The expected polarization degrees produced by the latter may reach, as in the case of \hbox{\object{Cyg~X-3}},
roughly $10$ and even $20$ per cent \citep[][where it is suggested that the polarization  is due to wind scattering]{2024NatAs...8.1031V}. This is comparable to the currently observed polarization degrees in bona fide XRPs but less than originally expected from such objects (see \citealt{2024Galax..12...46P} and references therein).
Most importantly, this polarization signal (especially the position angle) is going to vary with the spin phase.  
The polarization position angle is expected to be determined by the typical scattering plane of the beam of the rotating magnetosphere.
Pulsations in such `type-II ULXs' may be detectable with the new generation of X-ray polarization detectors (such as \emph{eXTP}, see \citealt{2019SCPMA..6229502Z}) even in objects where flux variations are beyond the detection limits. 
{As it is shown in \citet{2023AN....34420134P}, observations with \emph{eXTP} may produce high-quality spectra with the signal-to-noise sufficient for measuring the polarization degree as well as its spectral and spin-phase variations. }

\section{Summary}\label{sec:conc}

Increasing the mass accretion rate onto a strongly magnetized neutron star leads to a sequence of qualitatively different radiation-supported accretion regimes. 
We have shown that, when the expected shock height becomes comparable to or larger than the magnetospheric radius, a new regime becomes possible: optically thick, radiation-pressure-supported, shock-free magnetospheric flow. In this regime, the plasma remains subsonic throughout the magnetosphere.

Our main results are as follows.

\begin{enumerate}
    \item We derived a semi-analytical solution for a fully subsonic magnetospheric flow in the low-Mach-number limit. The solution is highly advective and has a nearly virial sound speed.

    \item Time-dependent one-dimensional simulations with {\tt HACol} reproduce the main properties of the analytical solution. The normalized enthalpy profile is close to the predicted one, while the energy density is regulated by mass and heat leakage through the sides of the flow.

    \item The luminosity produced by the column walls in this regime is limited by about several Eddington luminosities. Much higher luminosity is expected to be produced by the hot and mildly relativistic ejecta originating in the parts of the magnetosphere where force-free conditions are violated.

    \item  
    {A magnetosphere accreting in the shock-free regime is also likely to accrete from a disc violating its Eddington limit and launching a super-critical wind.} Scattering in the wind and the limited range of visible inclinations suppress the observed pulse fraction.

    \item We suggest that many non-pulsating ULXs may contain moderately magnetized neutron stars accreting in the shock-free regime. Pulsating ULXs may preferentially correspond to systems with stronger magnetic fields and larger magnetospheres.
\end{enumerate}

 \begin{acknowledgements} 
    This work was supported by the Deutsche
Forschungsgemeinschaft (DFG, German Research Foundation), project
number \mbox{570950648} (GL) and
a grant from the Simons Foundation (00001470, PA) and the International Space Science Institute (ISSI) in Bern, through ISSI International Team project \#495 (Feeding the spinning top). 
\end{acknowledgements}

\bibliographystyle{aa}
\bibliography{mybib}

\appendix

\section{Reduction to the Basko-Sunyaev's sinking solution}\label{app:BS}

In \citetalias{BS76}'s notation, $k = \frac{\gamma_{\rm BS}}{2\sin^2\theta_0}$, and $\sin^2\theta = \xi \sin^2\theta_0$. 
Neglecting higher-order terms in $\sin\theta \ll 1$, 
\begin{equation}
    e^{\pm k \cos \theta (1+\cos^2\theta)} \simeq e^{\pm \frac{\gamma_{\rm BS}}{\sin^2\theta_0}} e^{\mp\gamma\frac{\sin^2\theta}{\sin^2\theta_0}},
\end{equation}
that allows us to rewrite Eq.~(\ref{E:subsonic:f}) as
\begin{equation}
  \displaystyle  e^{-\gamma\xi} \frac{\diff}{\diff \xi} \left( e^{\gamma\xi} f\right) = \frac{3}{2\sin^2\theta_0} \frac{1}{\xi^2}.  
\end{equation}
This equation is integrable in special functions
\begin{equation}
   \displaystyle f = \frac{3}{4\sin^2\theta_0} e^{-\gamma\xi} \left[ \beta e^{-\gamma}  - E_2(\gamma) + \frac{1}{\xi} E_2(\gamma\xi)\right].
\end{equation}
Energy equation~(\ref{E:subsonic:energy}) is reduced in the  $\sin^2\theta \ll 1$ limit to 
\begin{equation}
    \frac{\diff}{\diff \xi} \ln u = - \frac{3}{\sin^2\theta_0} \frac{1}{\xi^2 f}.
\end{equation}
Substituting the analytic solution for $f$ yields
\begin{equation}
    \frac{\diff}{\diff \xi} \ln u = - 4 \frac{1}{\beta e^{-\gamma}  - E_2(\gamma) + \frac{1}{\xi} E_2(\gamma\xi)} \frac{e^{\gamma \xi}}{\xi^2}.
\end{equation}
The latter fraction in this expression is 
\begin{equation}
    \frac{e^{\gamma \xi}}{\xi^2} = - \frac{\diff }{\diff \xi} \left( \frac{1}{\xi} E_2(\gamma\xi)\right),
\end{equation}
that implies
\begin{equation}
    u \propto \left( \beta e^{-\gamma}  - E_2(\gamma) + \frac{1}{\xi} E_2(\gamma\xi)\right)^4.
\end{equation}
Applying the inner boundary condition reproduces equation (32) from \citetalias{BS76}.

\end{document}